\documentclass[conference]{IEEEtran}
\IEEEoverridecommandlockouts

\usepackage{cite}
\usepackage{amsmath,amssymb,amsfonts}
\usepackage{algpseudocode}
\usepackage{algorithm}
\usepackage{graphicx}
\usepackage{textcomp}
\usepackage{xcolor}
\usepackage{comment}
\usepackage{url}
\usepackage{subcaption}
 \usepackage{multirow}
\usepackage{amsmath}
\usepackage{amssymb}
\usepackage{booktabs}
\usepackage{hyperref}

\def\BibTeX{{\rm B\kern-.05em{\sc i\kern-.025em b}\kern-.08em
    T\kern-.1667em\lower.7ex\hbox{E}\kern-.125emX}}
\begin{document}

\title{Dynasto: Validity-Aware Dynamic–Static Parameter Optimization for Autonomous Driving Testing\\
} 

\author{
\IEEEauthorblockN{
Dmytro Humeniuk\IEEEauthorrefmark{1},
Mohammad Hamdaqa\IEEEauthorrefmark{1},
Houssem Ben Braiek\IEEEauthorrefmark{1},
Amel Bennaceur\IEEEauthorrefmark{4}
and
Foutse Khomh\IEEEauthorrefmark{1}
}

\IEEEauthorblockA{\IEEEauthorrefmark{1}
Polytechnique Montréal\\
Montréal, Canada\\
\{dmytro.humeniuk, mhamdaqa, houssem.ben-braiek, foutse.khomh \}@polymtl.ca
}
\IEEEauthorblockA{\IEEEauthorrefmark{4}
The Open University\\
Milton Keynes, UK\\
amel.bennaceur@open.ac.uk
}
}

\maketitle

\begin{abstract}
Extensive simulation-based testing is important for assuring the safety of autonomous driving systems (ADS). However, generating safety-critical traffic scenarios remains challenging because failures often arise from rare, complex interactions with surrounding vehicles. Existing automatic scenario-generation approaches frequently fail to distinguish genuine ADS faults from collisions caused by implausible or invalid adversarial behaviors, and they typically optimize either scenario initialization or agent behavior in isolation. We propose \textsc{Dynasto}, a two-step testing approach that jointly optimizes initial scenario parameters and dynamic adversarial behaviors to uncover realistic safety-critical failures. First, we train an adversarial agent using reinforcement learning (RL) with temporal-logic validity criteria and a safe-distance model, inspired by ISO~34502, to promote behaviorally plausible failures. Second, a genetic algorithm (GA) searches over initial conditions while replaying the adversary’s failure-inducing behaviors to reveal additional failures that the RL agent alone did not uncover. Finally, a graph-based clustering pipeline groups failures into representative modes based on semantic event sequences. Our evaluation experiments in \textit{HighwayEnv} across two ADS controllers show that \textsc{Dynasto} finds 60\%–70\% more valid failures than an RL-only adversary. With clustering, we obtain about 12 interpretable failure modes per system under test, revealing valid failures driven by weaknesses in ego-controller behavior. These results indicate that coordinated dynamic-static optimization with explicit validity constraints is effective for exposing safety-relevant failures in ADS testing.
\end{abstract}

\begin{IEEEkeywords}
Autonomous driving, Simulation-based testing, Scenario generation, Adversarial testing, Reinforcement learning, Genetic algorithms
\end{IEEEkeywords}

\section{Introduction}
Autonomous driving systems (ADS) must operate safely in highly dynamic traffic environments, where rare yet safety-critical situations emerge from complex interactions between vehicles~\cite{kalra2016driving}. Simulation-based testing is widely used to assess ADS behavior~\cite{birchler2025roadmap}, motivating a growing body of work on automatic scenario generation, where adversarial agents or search procedures construct traffic situations that expose weaknesses in the system under test (SUT). Early methods primarily searched over static initial conditions, such as the starting positions and speeds of traffic participants~\cite{abdessalem2018testing, calo2020generating}, or optimized pre-defined trajectories~\cite{ebadi2021efficient, huai2023doppelganger}. A key limitation is that the adversary’s behavior is fixed a priori and remains non-reactive to the SUT during execution, which reduces its ability to provoke realistic, interactive failures.

More recent approaches use reinforcement learning (RL) to learn adversarial driving policies or disturbance sequences that adapt online and steer the SUT toward failures~\cite{chen2021adversarial, haq2023many, doreste2024adversarial} by rewarding collision provoking behaviors. While this dynamic adaptation can substantially improve failure discovery, RL-based testing raises two persistent challenges addressed by \textsc{Dynasto}. First, it is difficult to design rewards that discourage reward-hacking behavior: adversaries may trigger collisions via unrealistic, overly aggressive maneuvers, and many existing testing methods treat all collisions as equally valuable~\cite{doreste2024adversarial, haq2023many}. This can obscure whether a collision reflects a genuine SUT brittleness, for example, failing to avoid a crash when a lead vehicle brakes from a safe initial headway, or an invalid interaction, for example, a cut-in at an unsafe distance that leaves the SUT insufficient time to react. Second, RL-based methods often operate under a fixed or narrowly sampled distribution of static scenario parameters (e.g., initial lanes, gaps, and speeds). Prior work in search-based testing~\cite{abdessalem2018testing, sorokin2024guiding} has shown that varying these initial conditions can be decisive for uncovering additional safety-critical failures.

In this work, we propose \textsc{Dynasto} (Validity-Aware Dynamic-Static Parameter Optimization for Autonomous Driving Testing), a two-step approach that jointly searches over dynamic adversarial behaviors and static scenario parameters and optimizes them in a coordinated way. In Step~1, we train an adversarial agent using RL to generate dynamic behaviors that challenge the ego vehicle (the SUT). The reward function is defined using temporal-logic-based validity criteria and a safe-distance model inspired by established safety standards (e.g., ISO~34502), so that only failures induced by unsafe ego behavior and respecting a minimum safe distance are promoted. Invalid failures due to unrealistically close or physically implausible maneuvers by the adversary are discarded. In Step~2, we run an evolutionary search over the initial conditions (e.g., initial positions, lanes, and speeds) of the failing scenarios identified in Step~1. Replaying the same failure-inducing behaviors under different static configurations allows revealing additional failures that were not found during the initial adversarial agent training phase.

To support post-hoc analysis, \textsc{Dynasto} includes a graph-based clustering procedure that groups failures into higher-level modes. Each failure trace is mapped to a sequence of discrete semantic events (e.g., cut-ins, cut-outs, and braking interactions), which is used to construct a $k$-nearest neighbors graph. We then apply the Leiden community detection algorithm~\cite{traag2019leiden} to identify clusters of failures that share similar interaction patterns, reducing hundreds of individual failures to a smaller number of representative patterns. We manually analyze the failures in the discovered clusters and highlight systematic weaknesses in the ego controller.

We evaluate \textsc{Dynasto} in the \textit{HighwayEnv} simulator~\cite{highway-env} on two systems under test (SUT1 and SUT2), corresponding to ego vehicles trained in two different environments. We further refer to environments with SUT1 as Use Case 1 (UC1) and those with SUT2 as Use Case 2 (UC2). \textit{HighwayEnv} is a 2D simulator commonly used in the RL community for self-driving policy training~\cite{dong2023comprehensive}. We compare our approach against random adversarial actions, a genetic algorithm (GA) that searches over fixed sequences of adversarial actions, the Deep Q-Learning~\cite{mnih2015human} baseline of Doreste et al.~\cite{doreste2024adversarial} that promotes all failures equally, and our own Deep Q-Network (DQN) adversary with a validity-aware reward. We further investigate how to combine dynamic action optimization with RL and search over static initial scenario parameters with GA. Specifically, we compare (i) co-evolutionary designs in which an RL adversary and a GA jointly optimize actions and initial conditions in parallel, and (ii) sequential designs in which the GA explores initial conditions only after the RL adversary has been trained and its failure-inducing behaviors can be replayed. Across these settings, \textsc{Dynasto} finds approximately 60\%–70\% more valid failures than a DQN-only adversary under the same evaluation budget and yields about 20\% more failure clusters than the co-evolutionary baseline. Finally, our post-hoc clustering analysis compresses roughly 200 unique failures per SUT into about 12 representative patterns.\\
In summary, this paper makes the following contributions:
\begin{itemize}
\item We introduce a validity semantics that separates genuine ADS faults from unrealistic crashes, via automatic trace labeling using Signal Temporal Logic (STL) rules aligned with ISO~34502 two-vehicle highway scenarios.
\item We present \textsc{Dynasto}, a two-step scenario-generation approach that couples RL-based adversarial action optimization with evolutionary search over initial conditions, guided by temporal-logic validity constraints.
\item We propose a graph-based clustering analysis that groups failures into interpretable modes using semantic event sequences, enabling scalable and structured failure analysis.
\item We evaluate two autonomous driving controllers in \textit{HighwayEnv}, comparing \textsc{Dynasto} against Random Search (RS), GA-only, and RL-based baselines, as well as co-evolutionary RL+Search Algorithm variants: GA or RS.
\item We analyze representative failure modes uncovered by \textsc{Dynasto}, showing how it exposes systematic weaknesses in ego-vehicle behavior.
\end{itemize}

\textbf{The remainder of this paper is organized as follows.} Section~\ref{sec:approach} presents \textsc{Dynasto}, including the validity criteria, RL formulation, GA configuration, and clustering pipeline. Section~\ref{sec:evaluation} describes the experimental setup and methodology, and reports quantitative and qualitative results on effectiveness and failure diversity. Section~\ref{sec:related} reviews related work. Section~\ref{sec:threats} discusses threats to validity, and Section~\ref{sec:conclusion} concludes the paper and outlines future directions.

\section{Approach}
\label{sec:approach}

\begin{figure*}[h]
\includegraphics[scale=0.35]{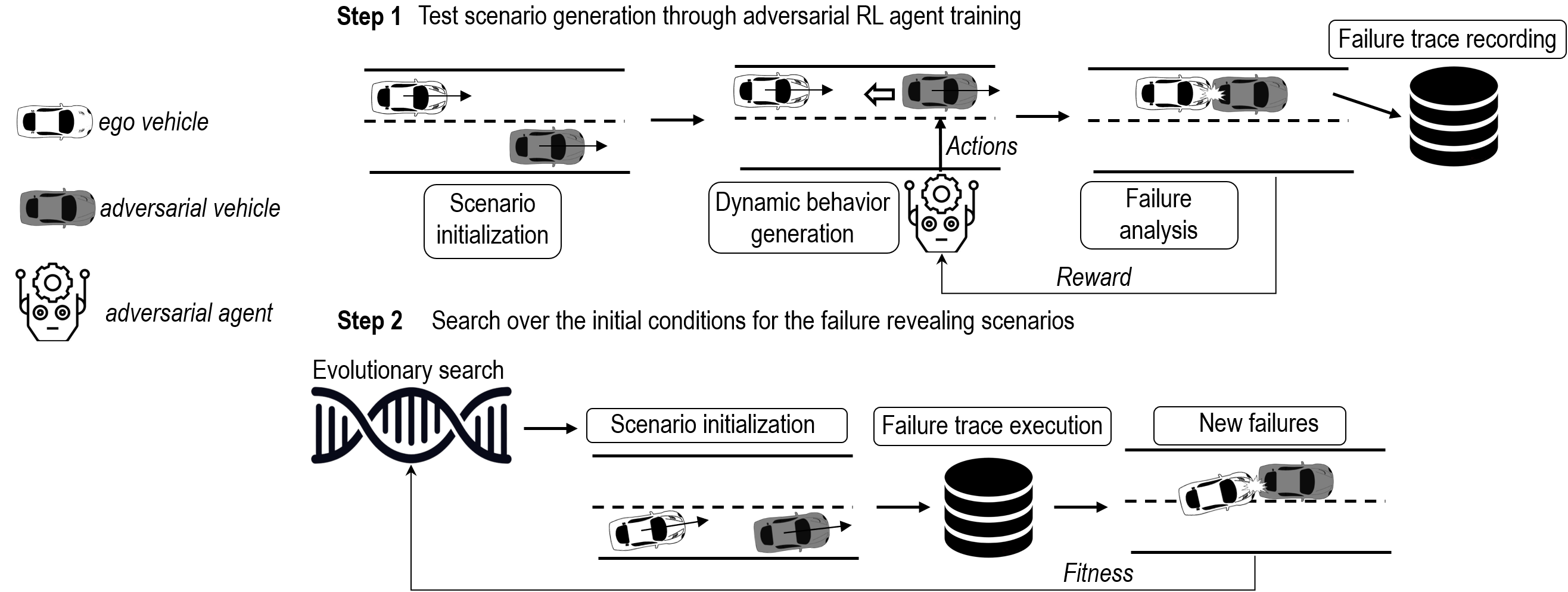}
\centering
\caption{The \textsc{Dynasto} Two-Step Approach Diagram}
\label{fig:Dynasto}
\end{figure*}

This section presents \textsc{Dynasto}, a validity-aware traffic scenario generation workflow that jointly optimizes \emph{dynamic adversarial behaviors} and \emph{static initial conditions} (Figure~\ref{fig:Dynasto}). \textsc{Dynasto} formalizes \emph{valid} failures with temporal-logic rules that filter out collisions caused by overly aggressive or unrealistic adversarial behavior, retaining only safety-relevant failures consistent with safe-distance constraints and scenario-specific interaction patterns. Step~1 trains an RL adversary to generate disturbance strategies that challenge the ego vehicle (SUT) while being rewarded only for valid failures, producing a set of failure traces with the adversary’s action sequences. Step~2 then searches over static initial parameters with a GA, evaluating each candidate by replaying the Step~1 adversary actions while the SUT responds online and re-checking validity under the same criteria. Finally, \textsc{Dynasto} enables post-hoc failure de-duplication and clustering to summarize failures into representative modes for analysis.


\subsection{Prerequisite: Valid Failure Modeling}
 \label{sec:failure-identification}
\textsc{Dynasto} relies on a precise failure definition. We target collisions that arise from behaviorally plausible, safety-relevant interactions, rather than from trivially unsafe or physically implausible trajectories. Central to this distinction is a \emph{safe-distance} requirement: the adversary must initiate its maneuvers with sufficient longitudinal headway so that, under normal conditions, the ego vehicle has enough time to react and avoid a collision. Collisions that occur despite this requirement are labeled \emph{valid} failures, while collisions triggered by violating the safe-distance condition are labeled \emph{invalid} and excluded from downstream analyses.
To bound the space of critical interactions, we follow ISO~34502~\cite{ISO34502}, which standardizes highway test scenarios for autonomous vehicles. The standard defines 24 critical scenarios, including 21 two-vehicle cases; we therefore prioritize two-vehicle scenarios. These scenarios span several zones, including the main highway, merge, and departure zones. In this work, we focus on two-vehicle scenarios in the main highway zone, leaving the remaining zones for future work. Our objective is to ensure that adversarial disturbances provide the SUT sufficient time to react and are not excessively aggressive. To this end, we focus on four ISO~34502 scenarios in which the adversary, also referred to as the principal other vehicle (POV), initiates the maneuver.

When a collision occurs, the trace is analyzed to determine whether the POV satisfied the corresponding safety constraints prior to the collision. If the constraints are respected, the failure is classified as valid; otherwise, it is considered invalid and attributed to overly aggressive POV behavior. We automate this trace classification using Signal Temporal Logic (STL), a common formalism for reasoning about cyber-physical system signals~\cite{tuncali2018simulation}. STL formulae are defined over predicates on signal variables and combine them using Boolean and temporal operators, including \emph{eventually} ($\Diamond_{\mathcal{I}}$), \emph{always} ($\Box_{\mathcal{I}}$), and \emph{until} ($\mathcal{U}_{\mathcal{I}}$), where the interval $\mathcal{I}$ specifies timing constraints.

\begin{figure}[h]
\includegraphics[scale=0.23]{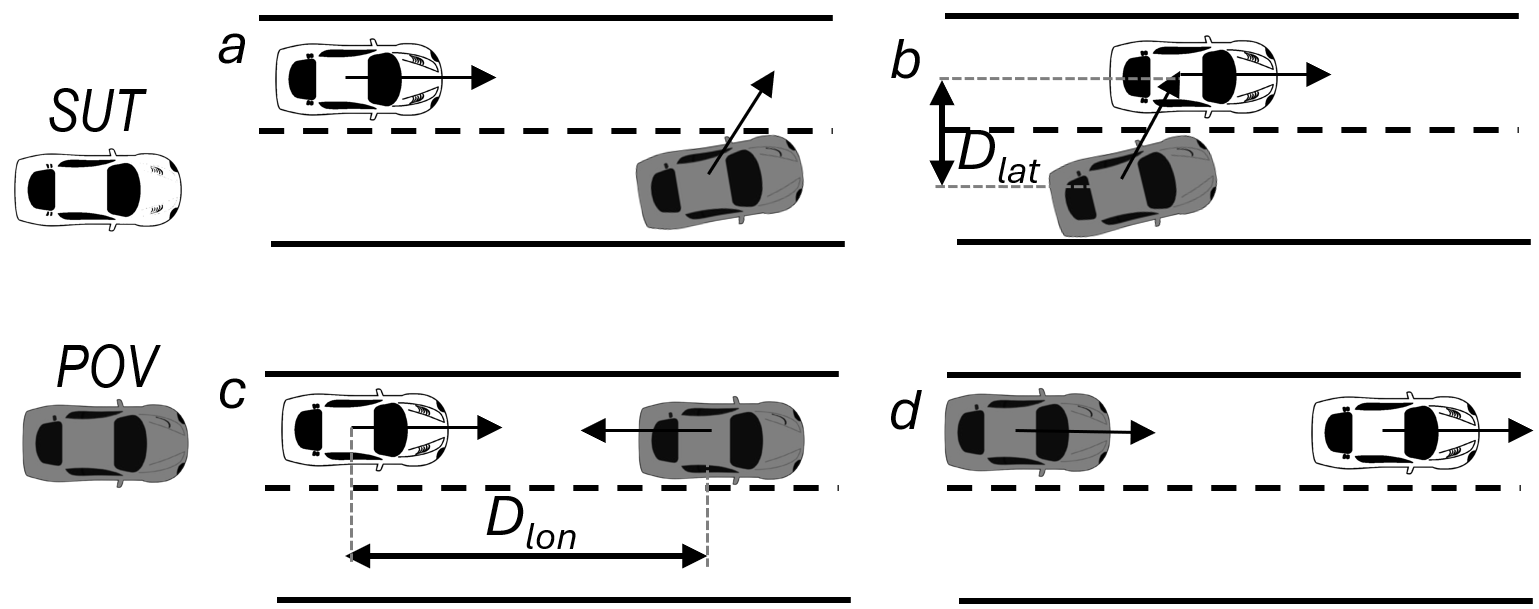}
\centering
\caption{Critical scenarios with maneuvers initialized by POV}
\label{fig:rules}
\end{figure}

The ISO~34502 scenarios selected for detailed analysis are shown in Figure~\ref{fig:rules}. Scenarios~$a$--$b$ correspond to collisions that occur after the POV changes lanes. In scenario~$a$, a failure is considered valid only if the safe longitudinal distance $D_{\text{lon}}$ is respected at the moment of the cut-in maneuver. In contrast, scenario~$b$ is always labeled invalid, since the required safe lateral distance $D_{\text{lat}}$ is violated during the cut-in. The definitions and computation of $D_{\text{lon}}$ and $D_{\text{lat}}$ are discussed in Section~\ref{sec:step1}. These distances specify the minimum admissible longitudinal and lateral separation required to ensure the following vehicle has sufficient time to react to the POV's maneuver.

To process the failure traces with STL rules, we first define a set of low-level predicates over the relative positions and accelerations of the ego (SUT) and adversary (POV) vehicles:

{\small
\begin{align*}
\mu_{\text{unsafe}}
:=\;& \big( d_{\text{lon}}(t_m) <  d_{\text{safe}}^{\text{lon}}(t_m) \wedge SameL \big)
\;\lor\;\\
&\qquad
   \big( d_{\text{lat}}(t_m) <  d_{\text{safe}}^{\text{lat}}(t_m) \wedge \neg SameL \big);\\
\mathit{Ahead_{adv}} :=\;& x_{\text{adv}}(t) > x_{\text{ego}}(t); \\
\mathit{LaneC_{adv}} :=\;& 
\bigl( \text{prev}(\text{lane}_{adv}) \neq \text{lane}_{adv} \bigr)
\wedge
\bigl( \text{prev}(\text{lane}_{ego}) = \text{lane}_{ego} \bigr);\\
\mathit{SameL} :=\;& 
\lvert y_{\text{adv}} - y_{\text{ego}} \rvert 
< \delta_{\text{lat}}, \;\text{same lane};\\
\mathit{Brake_{adv}} :=\;& a_{\text{adv}} \le -a_{\min}; \\
\mathit{Accel_{adv}} :=\;& a_{\text{adv}} \ge a_{\min}; \\
\mathit{CutIn_{adv}} :=\;&
\lozenge_{[t,t+\Delta t]}
\Big(\mathit{LaneC_{adv}}
    \wedge \mathit{Ahead_{adv}} \wedge SameL
\Big).
\end{align*}
}

Here $t_m$ corresponds to a timestep where one of the unsafe maneuvers shown in Figure \ref{fig:rules} $(a-c)$ was initiated, $\delta_{lat}$ a small threshold to determine if the two vehicles are in the same lane.
Firstly, the STL evaluation is applied only to traces that result in a collision, i.e., failures involving the two vehicles. Accordingly, each trace is analyzed over the interval $I \in [0, T_c]$, where $T_c$ refers to a timestep at which the collision occurs.

In the specified predicates, $\mu_{\text{unsafe}}$ detects whether the safe lateral or longitudinal distances are preserved, $\mathit{CutIn}$ determines if the cut-in maneuver was performed, $\mathit{Ahead}$ and $\mathit{Behind}$ whether the POV vehicle is ahead or behind the SUT, $\mathit{SameLane}$ detects if POV and SUT are in the same lane, $\mathit{Brake}$ and $\mathit{Accel}$ detect whether the POV is breaking or accelerating.

Based on the defined predicates, we formulate the following STL rules for the trace analysis: 
{\small
\begin{equation}
\varphi_{\text{unsafe-cut-in}}
:= \Diamond_{[0,\,T_c]}\!\big(
    \mathit{CutIn_{adv}}
    \wedge \mu_{\text{unsafe}}
\big)
\label{eq:unsafe-cut-in}
\end{equation}
\begin{equation}
\varphi_{\text{unsafe-brake}}
:= \Diamond_{[0,\,T_c]}\!\big(
    \mathit{Ahead_{adv}}
    \wedge  \mu_{\text{unsafe}}
    \wedge \mathit{Brake_{adv}} \wedge SameL
\big)
\label{eq:unsafe-brake-ahead}
\end{equation}
\begin{equation}
\varphi_{\text{rear-hit}}
:= \Diamond_{[0,\,T_c]}\!\Big(
    \mathit{Ahead_{ego}}
    \wedge \mathit{Accel_{adv}}\wedge SameL
\Big)
\label{eq:rear-hit}
\end{equation}
}

These STL specifications are evaluated on simulated traces to automatically detect and classify interactions. We use the \texttt{RTAMT} Python library~\cite{yamaguchi2024rtamt} to specify and evaluate the rules. If any rule evaluates to true, the collision is labeled \emph{invalid} and discarded; otherwise, it is labeled a \emph{valid} failure.

This labeling mechanism is important for the rest of \textsc{Dynasto}. In Step~1, it determines which outcomes contribute to the validity-aware reward used for adversarial training, and in Step~2, it provides the fitness signal for searching over initial conditions. As a result, both steps focus on behaviorally meaningful, safety-relevant failures rather than artifacts of unrealistic adversarial behavior.

\subsection{Step~1: Adversarial Policy Learning}
 \label{sec:step1}
In Step~1, \textsc{Dynasto} learns adversarial driving behaviors that trigger \emph{valid} failures as defined in Section~\ref{sec:failure-identification}. We cast the adversary as a reinforcement learning (RL) agent, encoding failures and validity into a scalar reward and updating the policy to maximize their occurrence. This formulation allows the adversary to generate temporally extended interaction patterns, such as sequences of lane changes and speed adjustments, conditioned on the current traffic context.

\subsubsection{Action space}
The adversarial agent operates in the discrete meta-action space provided by the \texttt{highway} environment in the \texttt{HighwayEnv} simulator. In this setting, the SUT drives on a two-lane highway with one adversarial vehicle. The SUT aims to maintain high speed while avoiding collisions. Actions are specified via the \texttt{DiscreteMetaAction} interface, which maps a small set of high-level driving intentions to predefined low-level vehicle controls. The action space contains five discrete actions:
$\texttt{LANE\_LEFT},\;\texttt{IDLE},\;\texttt{LANE\_RIGHT},\;\texttt{FASTER},\;\texttt{SLOWER}$.
These allow the adversary to change lanes (\texttt{LANE\_LEFT}, \texttt{LANE\_RIGHT}), adjust longitudinal speed (\texttt{FASTER}, \texttt{SLOWER}), or maintain its current motion (\texttt{IDLE}). This abstraction hides continuous vehicle dynamics, enabling efficient policy learning while preserving the key behavioral primitives needed to induce realistic interactions.

\subsubsection{Reward Formulation}
Our adversarial-training reward induces collisions while prioritizing behaviorally plausible, safety-relevant failures. It combines (i) a collision-likelihood shaping term based on the distance between the ego vehicle and the adversary and (ii) a terminal bonus awarded only when the resulting collision is classified as \emph{valid} under the temporal-logic criteria in Section~\ref{sec:failure-identification}. Distance-based collision likelihood is commonly used in prior work and is adopted in approaches such as \textit{DeepCollision}~\cite{lu2022learning}.

\paragraph{Safe-distance model}
Let $v_{\mathrm{l}}$ and $v_{\mathrm{f}}$ denote the longitudinal velocities of the leading and following vehicles, respectively. Both vehicles are in the same lane, with the adversary leading and the SUT following. The safe longitudinal distance $d_{\text{safe}}^{\text{lon}}$ is computed using a constant-deceleration model based on a Berkley algorithm~\cite{seiler1998development}, which is commonly used to compute safe distance~\cite{lu2022learning}:
\begin{equation}
    d_{\text{safe}}^{\text{lon}}(t)
        = 
        \frac{1}{2}
        \Bigg(
            \frac{v_{\mathrm{l}}(t)^{2}}{ a_{\mathrm{l}}}
            \;-\;
            \frac{v_{\mathrm{f}}(t)^{2}}{ a_{\mathrm{f}}}
        \Bigg)
        \;+\;
        v_{\mathrm{f}}(t) \cdot \tau_{\mathrm{r}}
        \;+\;
        d_{\text{min}}^{\text{lon}} .
\end{equation}
Here, $a_{\mathrm{l}}$ and $a_{\mathrm{f}}$ denote the maximum decelerations of the lead and following vehicles. In our experiments, we use the default values from \texttt{HighwayEnv}, setting both to $-5\,\mathrm{m/s^2}$. The minimum allowable longitudinal distance $d_{\text{min}}^{\text{lon}}$ is set to $5\,\mathrm{m}$ following \textit{DeepCollision}. Unlike \textit{DeepCollision}, which assumes $\tau_r = 0\,\mathrm{s}$ for an autonomous SUT, we set $\tau_r = 0.2\,\mathrm{s}$, motivated by reports indicating that expected ADS reaction times are on the order of $0.1\,\mathrm{s}$~\cite{entrapeer2023autonomous}.

When the vehicles are in different lanes, we use a safe lateral distance $d_{\mathrm{safe}}^{\mathrm{lat}}$, computed based on~\cite{reimann2024temporal}:
\begin{equation}
d_{\mathrm{safe}}^{\mathrm{lat}}(t)
    =
    v_{lat}(t)\tau_r
    + \frac{1}{2} a_{\max}^{\mathrm{lat}} \tau_r^2
    + \frac{\bigl( v_{lat}(t) + a_{lat}\tau_r \bigr)^2}
           {2 b_{lat} + d_{\text{min}}^{\text{lat}}} .
\end{equation}
Here, $v_{\mathrm{lat}}$ denotes the lateral velocity of the adversarial vehicle, $\tau_r$ is the reaction time, and $a_{\mathrm{lat}}$ and $b_{\mathrm{lat}}$ represent the lateral acceleration and deceleration, respectively. We set $\tau_r = 0.2\,\mathrm{s}$, $a_{\mathrm{lat}} = 5\,\mathrm{m/s^2}$, $b_{\mathrm{lat}} = -5\,\mathrm{m/s^2}$, and $d_{\text{min}}^{\text{lat}} = 2\,\mathrm{m}$.

At each timestep $t$, we compute the current distance between the vehicles as
\begin{equation}
d_{c}(t) = \sqrt{(x_{\mathrm{ego}}(t) - x_{\mathrm{adv}}(t))^{2} + (y_{\mathrm{ego}}(t) - y_{\mathrm{adv}}(t))^{2}} .
\end{equation}

The required safe distance is then
{\small
\begin{equation*}
d_{\mathrm{safe}}(t) =
\begin{cases}
d_{\mathrm{safe}}^{\mathrm{lon}}(t), & \text{if the vehicles are in the same lane}, \\
d_{\mathrm{safe}}^{\mathrm{lat}}(t), & \text{if the vehicles are in different lanes}.
\end{cases}
\end{equation*}
}

\paragraph{Collision-likelihood shaping term}
Let $d_{\mathrm{c}}(t)$ be the current distance between the two vehicles. We define the collision-likelihood reward as:
{\small
\begin{equation}
    R_{\mathrm{coll}}(t) =
    \begin{cases}
        \displaystyle
        \frac{d_{\mathrm{safe}}(t) - d_{\mathrm{c}}(t)}{d_{\mathrm{safe}}}, 
        & \text{if } d_{\mathrm{c}}(t) < d_{\mathrm{safe}}(t), \\[10pt]
        0, & \text{otherwise}.
    \end{cases}
\end{equation}
}
This term is assigned at each simulation step and increases as the adversary reduces the distance below the safe threshold, encouraging situations in which a collision becomes likely.

\paragraph{Validity-aware terminal bonus}
{\small
\begin{equation}
    R_{\mathrm{valid}} =
    \begin{cases}
        \displaystyle
        30, 
        & \text{if failure is valid}, \\[10pt]
        \displaystyle
        -\frac{d_{\mathrm{safe}}(t_m) - d_{c}(t_m)}{5}, 
        & \text{otherwise}.
    \end{cases}
\end{equation}
}
If the collision is invalid, we compute $d_{\mathrm{safe}}(t_m)$ and $d_c(t_m)$ at the timestep $t_m$ when the adversary initiated the unsafe maneuver and penalize violations of the safe-distance requirement. This term is applied once per episode, after a collision occurs.

\paragraph{Total reward and episode termination}
The full reward is
\begin{equation}
R = \sum_{t=0}^{T} R_{\mathrm{coll}}(t) + R_{\mathrm{valid}} .
\label{eq:full_eq}
\end{equation}
The maximum episode duration is $T=40$ steps. An episode terminates earlier at step $T_c$ if a collision occurs. In this way, the adversary is encouraged not only to approach the ego vehicle dangerously, but to synthesize interactions that qualify as valid failures under our logical specifications.

\subsubsection{State Representation.}
The adversary observes the positions and velocities of both vehicles, as summarized in Table~\ref{tab:state_repr}.

\begin{table}[h!]
\centering
\caption{State representation provided to the adversarial agent.}
\begin{tabular}{lcccc}
\toprule
\textbf{Vehicle} & $x$ & $y$ & $v_x$ & $v_y$ \\
\midrule
ego-vehicle & $x_{\mathrm{ego}}$ & $y_{\mathrm{ego}}$ & $v_{x,\mathrm{ego}}$ & $v_{y,\mathrm{ego}}$ \\
adversary   & $x_{\mathrm{adv}}$ & $y_{\mathrm{adv}}$ & $v_{x,\mathrm{adv}}$ & $v_{y,\mathrm{adv}}$ \\
\bottomrule
\end{tabular}
\label{tab:state_repr}
\end{table}

These features allow the policy to infer relative position, lane configuration, and closing speed, which are essential for planning lane changes and speed adjustments. The SUT observes the same state.

\subsubsection{Adversarial RL training and Step~1 outcomes}
We train the RL adversary using \textit{DQN}~\cite{mnih2015human} implemented in \texttt{CleanRL}~\cite{huang2022cleanrl}. We also evaluated \textit{PPO}~\cite{schulman2017ppo} and \textit{SAC}~\cite{haarnoja2018sac}, but \textit{DQN} consistently achieved the strongest results, and we therefore adopt it in \textsc{Dynasto}. Combined with the state representation, meta-action space, and reward formulation, this training yields an adversarial policy that reliably induces \emph{valid} failures in the ego controller under the validity constraints of Section~\ref{sec:failure-identification}. Training produces a repertoire of failure-inducing behaviors for a fixed family of initial conditions. For each valid failure, we record the associated initial conditions and the adversary action at each timestep. These traces form the dynamic component that Step~2 replays while searching over initial scenario parameters to uncover additional failures driven by the interaction between static initialization and adversarial behavior.

\subsection{Step~2: Initial Condition Search}
\label{sec:step2}

In Step~2, \textsc{Dynasto} targets the \emph{static} initial conditions, such as vehicle positions, lanes, and speeds, which are fixed at episode start and lie outside the RL action space. Although Step~1 optimizes the dynamic action sequence, these parameters shape the resulting trajectory and determine whether replaying an adversarial behavior yields a valid failure. Step~2 therefore broadens the exploration beyond the relatively narrow initial-condition distribution induced by the default \texttt{highway} settings, complementing the stability-driven training regime of Step~1.

\subsubsection{Black-box formulation and search objective}
After an episode terminates, we treat the rollout as a black-box evaluation of a static initialization. For a given set of initial parameters, replaying the adversarial actions recorded in Step~1 yields a trace, and the failure-identification rules from Section~\ref{sec:failure-identification} determine whether it contains a valid failure. This motivates a metaheuristic search over static initial conditions, where each candidate is evaluated by a single rollout and the resulting outcome serves as a fitness signal. We adopt a GA because it is gradient-free, supports mixed continuous and discrete parameters, and can exploit episodic fitness derived from valid failures.

\subsubsection{Search algorithm implementation}
To search over the space of initial scenario parameters, we used the GA implementation from Pymoo library \cite{pymoo} with a continuous, normalized genotype representation. The GA evolves a population of candidate initial configurations, with each individual encoded as a vector in $[0,1]^d$. The search objective is to discover initialization settings that induce failures when the failing scenarios recorded in Step~1 are rolled out.

\paragraph{Individual representation}
Each GA individual encodes a complete initial configuration of the ego and adversary vehicles. Let $\mathbf{x} \in [0,1]^d$ be a genotype vector of dimension $d=11$, corresponding to the parameters (position, current lane, heading, speed, target lane and id of selected scenario from Step~1) of both vehicles. The normalization is parameter-wise:
\[
x_i = \frac{p_i - p_i^{\min}}{p_i^{\max} - p_i^{\min}},
\]
where $p_i$ is the raw scenario parameter and $(p_i^{\min}, p_i^{\max})$ are its admissible bounds. 
During phenotype reconstruction, the inverse transformation
\[
p_i = p_i^{\min} + x_i (p_i^{\max} - p_i^{\min})
\]
is applied, followed by rounding when $p_i$ corresponds to a discrete parameter such as a lane identifier or scenario id. 

This individual representation enables the GA to optimize continuous and discrete scenario attributes within a single real-valued search space.

\paragraph{Initial population}
The initial population is generated by independently sampling each parameter within its admissible range. For each dimension $i$, a value $p_i$ is drawn from either a uniform real interval or an integer range, depending on its type. The resulting parameter vector is then normalized to $[0,1]^d$ using the above-described transformation. All individuals share the same fixed-length encoding.

\paragraph{Crossover}
The GA uses Simulated Binary Crossover (SBX) \cite{deb2007self} on pairs of parent vectors to generate offspring. SBX is well suited to real-valued encodings and mimics the behavior of single-point crossover on binary strings, producing offspring distributed around the parents according to a polynomial probability distribution. SBX ensures the exploitation of promising generated pairs.

\paragraph{Mutation}
After crossover, polynomial mutation (PM) \cite{deb2014analysing} is applied independently to each dimension of an offspring vector. This operator perturbs each component with a probability given by the mutation rate, yielding small but effective local modifications while keeping all values within the $[0,1]$ bounds. PM promotes the exploration of the parameter space while evolving the best ones. 

\paragraph{Fitness function}
We rely on Equation~\ref{eq:full_eq} to evaluate the fitness of individuals, analogous to the reward calculation used for the adversary RL agent. This fitness value is assigned after the entire scenario has been executed and is multiplied by a negative sign, since the \emph{Pymoo} framework supports only minimization objectives.

\paragraph{Duplicate removal} 
We promote population diversity by discarding redundant individuals whose genotype matches one already present in the evaluated population. Using pairwise Euclidean distance with threshold $D_{th}$, two individuals are considered distinct only if they differ by at least 1.5\% per dimension on average.

\subsubsection{Simulation-based evaluation and Step~2 outcome}
Unlike RL-based dynamic optimization, where agents interact with \texttt{HighwayEnv} at every simulation step, GA-based static optimization searches over initial conditions, with each individual representing a complete test configuration. Fitness evaluation therefore requires executing a full simulation episode. To integrate this process into the existing RL scenario-execution pipeline, we use an \emph{ask-tell} interface. In the \textit{ask} phase, the GA outputs a candidate test (a vector of initial positions, headings, speeds, and lane indices) without running any simulation. The candidate is injected into the environment and rolled out using the SUT policy and the recorded adversary actions. After termination, the resulting fitness is returned to the GA in the \textit{tell} phase, and the process repeats for each individual across generations. This evaluation loop drives Step~2. Each candidate is decoded into a concrete initialization and paired with a Step~1 scenario selected via the decoded id. The recorded adversary actions provide the dynamic component, while the SUT continues to act online, as in Step~1. The resulting trace is assessed using the failure-identification rules from Section~\ref{sec:failure-identification}, and configurations that trigger valid failures receive higher fitness and are preferentially propagated. Consequently, Step~2 expands the failure set beyond those seen during RL training by systematically varying static initial parameters while replaying failure-inducing behaviors. Step~2 is agnostic to how Step~1 produces failures; we use an RL-based adversary because it outperformed baselines in our evaluation.

\subsection{Post-hoc: Failure De-duplication and Analysis}
\label{sec:post_hoc_analysis}
The two-step scenario generation can produce a large set of valid failures, many of which are redundant or minor variations of the same underlying behavioral pattern. We therefore perform post-hoc failure de-duplication and analysis to remove redundancy and to assess how different testing configurations explore the failure space and its diversity.

\subsubsection{Failure De-duplication}
During scenario generation, all discovered scenarios are retained in memory. Afterward, highly similar recorded failures are grouped and filtered as duplicates, so they are not counted as distinct, unique failures. To compare failures, we represent each scenario with a descriptive vector that concatenates the normalized relative positions and speeds of the ego vehicle and the adversary over the last $N$  steps before collision, yielding a $3 \times N$-dimensional embedding. Empirically, a value of $N=8$ was chosen. Similarity is measured using a Euclidean-distance threshold $S_{th}$, calibrated on a manually curated set of about 100 similar and distinct scenarios. This calibration set is included in the replication package.

\subsubsection{Failure Mode Clustering}
\begin{figure*}[h]
\includegraphics[scale=0.3]{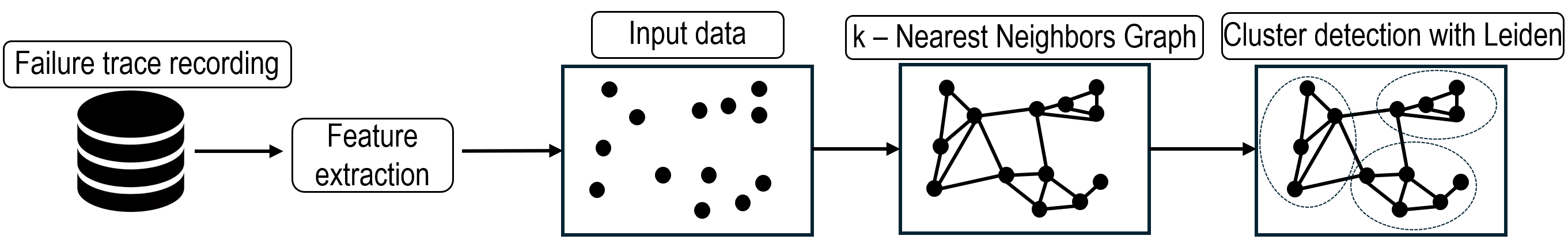}
\centering
\caption{Failure cluster detection approach}
\label{fig:Dynasto-clustering}
\end{figure*}

As illustrated in Figure~\ref{fig:Dynasto-clustering}, we apply a graph-based clustering pipeline that groups similar failure traces into coherent failure modes using the next steps.

\paragraph{Failure trace recording}
After a valid failure is identified, we record its full failure trace, including the time series of system states and the actions of both the adversarial and ego agents. We then post-process this trace using STL rules to detect events such as cut-ins, cut-outs, and braking by either the adversary or the ego vehicle. To support this event detection, the STL rules introduced in Section~\ref{sec:failure-identification} are extended accordingly.

\paragraph{Feature extraction}
In Section~\ref{sec:post_hoc_analysis}, we introduce \emph{descriptive vectors} that extract relative positions and speeds from each scenario trace. While these features capture low-level geometric and kinematic properties of the interaction, they often lack semantic structure and can be sensitive to minor trajectory variations. As a result, failures that are similar at a higher level, e.g., two cut-in scenarios with slightly different lateral offsets, may appear distinct under a purely low-level representation. For clustering analysis, we therefore extract higher-level features with clearer semantic meaning.\\
For each failure trace, we extract a sequence of high-level semantic events that characterizes how the interaction between the ego vehicle and the adversarial vehicle evolves over time. We use discrete behavioral descriptors such as lane-change patterns and braking interactions. To detect these events, we complement the STL rules from Section~\ref{sec:failure-identification} (used to identify valid failures) with the following predicates:

{\footnotesize
\begin{align*}
\mathit{Side_{adv}} :=\;& x_{\text{ego}}(t) - V_L < x_{\text{adv}}(t) < x_{\text{ego}}(t) + V_L; \\
\mathit{Side_{ego}} :=\;& x_{\text{adv}}(t) - V_L < x_{\text{ego}}(t) < x_{\text{adv}}(t) + V_L; \\
\mathit{Brake_{ego}} :=\;& a_{\text{ego}} \le -a_{\min};\\
\mathit{Accel_{ego}} :=\;& a_{\text{ego}} \ge a_{\min};\\
\mathit{LaneC_{ego}} :=\;& 
\bigl( \text{prev}(\text{lane}_{ego}) \neq \text{lane}_{ego} \bigr)
\wedge
\bigl( \text{prev}(\text{lane}_{adv}) = \text{lane}_{adv}\bigr);
\\
\mathit{CutIn_{ego}} :=\;&
\lozenge_{[t,t+\Delta t]}
\Big(\mathit{LaneC_{ego}}
    \wedge \mathit{Ahead_{ego}} \wedge SameL
\Big);\\
\mathit{CutOut_{adv}} :=\;&
\lozenge_{[t,t+\Delta t]}
\Big(\mathit{LaneC_{adv}}
    \wedge \mathit{Ahead_{adv}} \wedge \neg SameL
\Big);\\
\mathit{CutOut_{ego}} :=\;&
\lozenge_{[t,t+\Delta t]}
\Big(\mathit{LaneC_{ego}}
    \wedge \mathit{Ahead_{ego}} \wedge \neg SameL
\Big);\\
\mathit{CutInSide_{ego}} :=\;&
\lozenge_{[t,t+\Delta t]}
\Big(
   \mathit{LaneC_{ego}}
    \wedge SameL
\;    \wedge \mathit{Side_{ego}}
\Big);\\
\mathit{CutInSide_{adv}} :=\;&
\lozenge_{[t,t+\Delta t]}
\Big(
    \mathit{LaneC_{adv}}
\;\wedge \mathit{SameL} 
    \wedge \mathit{Side_{adv}}
\Big);\\
\mathit{BrakeSL_{adv}}
:=\;& \lozenge_{[t,t+\Delta t]}\!\big(
    \mathit{Ahead_{adv}}
    \wedge \mathit{Brake_{adv}} \wedge SameL
\big);\\
\mathit{BrakeDL_{adv}}
:=\;& \lozenge_{[t,t+\Delta t]}\!\big(
    \mathit{Ahead_{adv}}
    \wedge \mathit{Brake_{adv}} \wedge \neg SameL.
\big)
\end{align*}
}
Here $\mathit{Side_{adv}}$ and $\mathit{Side_{ego}}$ identify side-by-side configurations, where $V_L$ is the vehicle length. These conditions are needed to identify side cut-ins. The predicates $\mathit{Brake_{ego}}$ and $\mathit{Accel_{ego}}$ capture braking and acceleration events of the ego vehicle based on a minimum acceleration threshold $a_{\min}$. Lane-change events are encoded using $\mathit{LaneC_{ego}}$, which evaluates whether the ego vehicle has switched lanes relative to the previous timestep (denoted by $prev()$) while the adversary remained in its lane. $\mathit{SameL}$ predicate is defined in Section  \ref{sec:failure-identification} and corresponds to both vehicles remaining in the same lane. $\mathit{CutIn_{ego}}$ identifies an ego cut-in maneuver, while $\mathit{CutInSide_{ego}}$ and $\mathit{CutInSide_{adv}}$ detect lateral cut-in events in which a lane change leads to a side-by-side configuration. $\mathit{CutOut_{adv}}$ characterizes an adversary cut-out, where the adversary changes lanes and moves outside the ego's lane. Finally, $\mathit{BrakeSL_{adv}}$ and $\mathit{BrakeDL_{adv}}$ describe adversary braking events in the same lane or a different lane, respectively.\\
These predicates are evaluated at each simulation step and mapped to an integer identifier according to the following taxonomy:

{\small
\begin{itemize}
    \item[0:] none of the events below occurred;
    \item[1:] \texttt{CutInSideEgo}, the ego changes lane while parallel to the adversary and collides with it;
    \item[2:] \texttt{CutInEgo}, the ego cuts in ahead of the adversary;
    \item[3:] \texttt{CutOutEgo}, the ego cuts out in front of the adversary;
    \item[4:] \texttt{CutInSideAdv}, the adversary changes lane while parallel to the ego and collides with it;
    \item[5:] \texttt{CutOutAdv}, the adversary cuts out of the lane ahead of the ego;
    \item[6:] \texttt{CutInAdv}, the adversary cuts in ahead of the ego;
    \item[7:] \texttt{BrakeSameLaneAdv}, the adversary slows down in front of the ego in the same lane;
    \item[8:] \texttt{BrakeDiffentLaneAdv}, the adversary slows down in front of the ego in a different lane;
    \item[999:] collision event.
\end{itemize}
}

Each failure trace is then represented as a time-ordered vector of event identifiers, which we refer to as an \emph{event vector}. For example, the vector $[0,0,0,6,0,0,7,7,999]$ denotes a sequence in which the adversary first performs a cut-in maneuver, then slows down in the same lane, and ultimately a collision occurs. If multiple events occur at the same time step, we record their sum. A subset of events (0--6) is mutually exclusive, ensuring that no ambiguity arises when using the sum-based encoding. To compare event sequences, we use the Levenshtein distance~\cite{navarro2001guided}, which measures the minimum number of edit operations (insertions, deletions, and substitutions) required to transform one sequence into another. Thus, similarity depends on how many high-level events differ rather than on the absolute numerical values of the identifiers.

\paragraph{Construction of the $k$--nearest neighbors graph}
Given the matrix of event vectors, we compute pairwise distances using the selected distance metric (Levenshtein distance). For each failure trace, we identify its $k$ most similar neighbors and connect them with undirected edges, forming a $k$--nearest neighbors ($k$NN) graph. This graph captures the local similarity structure of the failure space: highly similar traces become densely connected, whereas outliers form sparse local structures. We assign edge weights proportional to similarity (i.e., $w = 1 / (1 + d)$, where $d$ is the distance between two traces), which guides the subsequent step.


\paragraph{Cluster detection with Leiden}
To automatically group similar failure traces into coherent modes, we apply the Leiden community-detection algorithm~\cite{traag2019leiden} to the weighted $k$NN graph. Leiden optimizes the modularity objective~\cite{newman2006modularity}, favoring partitions with dense within-community connectivity and sparse connectivity across communities, and typically yields stable, well-separated clusters. Under this view, a failure mode corresponds to a community of traces that are strongly connected to each other but only weakly connected to the rest of the graph, capturing a shared underlying behavioral pattern. Our main motivation for using a community-detection algorithm is twofold. First, Leiden assigns every point to a cluster and does not rely solely on distance thresholds. Instead, it leverages the connectivity structure of the $k$NN graph to extract well-connected communities, whereas density-based methods often yield many outliers that require manual inspection. Second, Leiden is simple to tune, requiring only a small set of parameters, mainly the number of nearest neighbors (NN) and the distance metric. This makes it more practical than alternatives such as DBSCAN, HDBSCAN, or hierarchical clustering, which are commonly used in the testing literature but can be highly sensitive to hyperparameter choices \cite{attaoui2024supporting}. In our cases, we selected the number of nearest neighbors (NN) using a labeled set of about 100 scenarios organized into seven clusters. We then tuned the distance metric so that the clusters produced by Leiden align with this reference set.
\paragraph{Clustering analysis outcome}
The resulting clusters correspond to distinct categories of failure modes (e.g., cut-ins, late braking, lane-change collisions). This clustering facilitates reasoning about the coverage and diversity of the revealed failures, supports the identification of dominant patterns, and highlights rare but safety-critical behaviors. It also provides a structured basis for comparing different testing configurations in terms of the failure modes they expose.

\section{Evaluation}
\label{sec:evaluation}
In this section, we first describe the experimental setup, then report quantitative and qualitative results.

\subsection{Experimental setup}
Below is our experimental setup for empirical evaluation.\\
\textbf{Evaluation budget and statistical protocol.} For each evaluated method, we allocate a total budget of 4000 steps. For \textsc{Dynasto}, we assign 3000 steps to RL training in Step~1 and 1000 steps to the initial-condition search in Step~2. We repeat each evaluation at least 10 times. We report the non-parametric Mann-Whitney U test with significance level $\alpha = 0.05$, along with Cliff's delta effect size~\cite{mann1947test,macbeth2011cliff}. The results of the statistical testing are illustrated in the boxplots. Pairwise statistical comparisons between configurations are indicated above the boxplots using horizontal brackets. Statistical significance is denoted using a star-based notation, where ``**'' indicates a statistically significant difference ($p < 0.01$) and ``*'' indicates ($p < 0.05$). Effect sizes are reported using dot notation, where ``..'' denotes a large effect and ``.'' denotes a medium effect size.\\
\textbf{Compute.} All experiments are run on a PC with a 16-core AMD Ryzen 7 4800HS CPU @ 2.90 GHz, 16 GB of memory, and an NVidia GeForce GTX 1660 GPU with 6 GB.\\
\textbf{Step~1 configuration.} The RL adversary is trained with learning rate $5 \times 10^{-4}$, batch size 64, $\gamma=0.95$, and a replay buffer of 150{,}000, using $\varepsilon$-greedy exploration with $\varepsilon$ decaying from 1.0 to 0.05 over the first 20\% of training. Initial conditions use two lanes, speed 25~m/s, heading 0~rad, with $x_{\text{ego}}!\sim[247,263]$ and $x_{\text{adv}}!\sim[295,327]$.\\
\textbf{Step~2 configuration.} We use a GA with population size 100 and crossover and mutation probabilities of 0.9. Parameters are sampled from Table~\ref{tab:init_param_ga} and include $(x_{\text{ego}},x_{\text{adv}})$, $(h_{\text{ego}},h_{\text{adv}})$, $(l_{\text{ego}},l_{\text{adv}})$, and $(tl_{\text{ego}},tl_{\text{adv}})$ (0 and 1 denote left and right lanes), plus a failure-id variable indexing recorded traces. We set the duplicate-removal threshold to $D_{th}=0.05$.

\begin{table}[h!]
\centering
\caption{Initial parameter ranges for Ego and Adversary.}
\scalebox{0.85}{
\begin{tabular}{ccccccccccc}
\hline
\textbf{Value} & \textit{$x_{\text{ego}}$} & \textit{$x_{\text{adv}}$} & \textit{$l_{\text{ego}}$} & \textit{$l_{\text{adv}}$} & \textit{$tl_{\text{ego}}$} & \textit{$tl_{\text{adv}}$} & \textit{$h_{\text{ego}}$} & \textit{$h_{\text{adv}}$} & \textit{$s_{\text{ego}}$} & \textit{$s_{\text{adv}}$} \\
\hline
Min & 247 & 395 & 0 & 0 & 0 & 0 & -0.08 & -0.08 & 20 & 20 \\
Max & 304 & 364 & 1 & 1 & 1 & 1 & 0.08 & 0.08 & 29 & 29 \\
\hline
\end{tabular}}
\label{tab:init_param_ga}
\end{table}

\noindent \textbf{Systems under test.} We evaluate two ego controllers trained with DQN in the \texttt{highway} environment of \texttt{HighwayEnv}, reusing the SUTs from Doreste et al.~\cite{doreste2024adversarial} for transferability and baseline comparison. Surrounding vehicles follow IDM policies~\cite{treiber2000congested}: SUT1 is trained with one standard IDM vehicle, and SUT2 with four defensive IDM vehicles. Both are trained for 8000 steps with rewards for high speed, collision avoidance, and staying in the right lane~\cite{doreste2024adversarial}.

\subsection{Results}
Below, we report the results organized in 3 research questions. 

\subsubsection{RQ1 -- Effectiveness of \textsc{Dynasto} against Baselines}
\hfill \break
\noindent \textbf{Motivation.} In this RQ, we evaluate the effectiveness of \textsc{Dynasto} against common adversarial test-generation baselines using the number of \emph{valid, unique} failures uncovered under a fixed testing budget. \\
\textbf{Method.} We compare \textsc{Dynasto} against four baselines on the two systems under test, SUT1 (UC1) and SUT2 (UC2):
\begin{itemize}
 \item \textit{Random Search (RS)}: at each step, the adversarial action is sampled uniformly from the available action set.
 \item \textit{Genetic Algorithm (GA)}: GA precomputes a fixed adversarial action sequence before execution, yielding open-loop behavior that cannot adapt online to the ego.
 \item \textit{Baseline (Base) (Doreste et al.~\cite{doreste2024adversarial})}: RL adversary trained with Deep Q-Learning that selects actions online but rewards all failures, without separating valid from invalid ones.
 \item \textit{Validity-Aware RL (VARL)}: our RL adversary, whose reward explicitly promotes only valid failures under the temporal-logic criteria in Section~\ref{sec:failure-identification}.
 \end{itemize}
 
\begin{figure}[h!]
\includegraphics[scale=0.35]{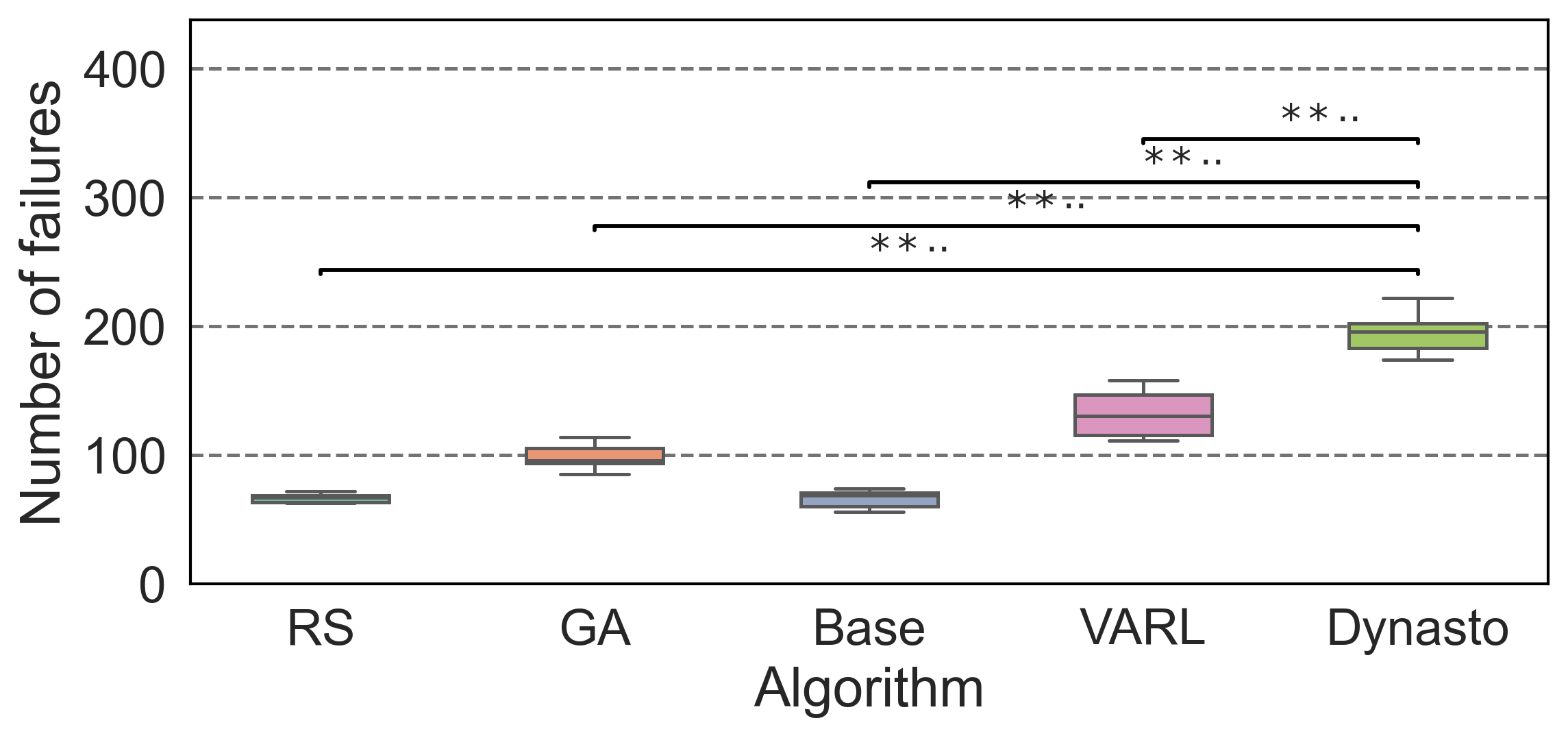}
\centering
\caption{Number of failures revealed by algorithm in UC1}
\label{fig:uc1_box}
\end{figure}

\begin{figure}[h!]
\includegraphics[scale=0.35]{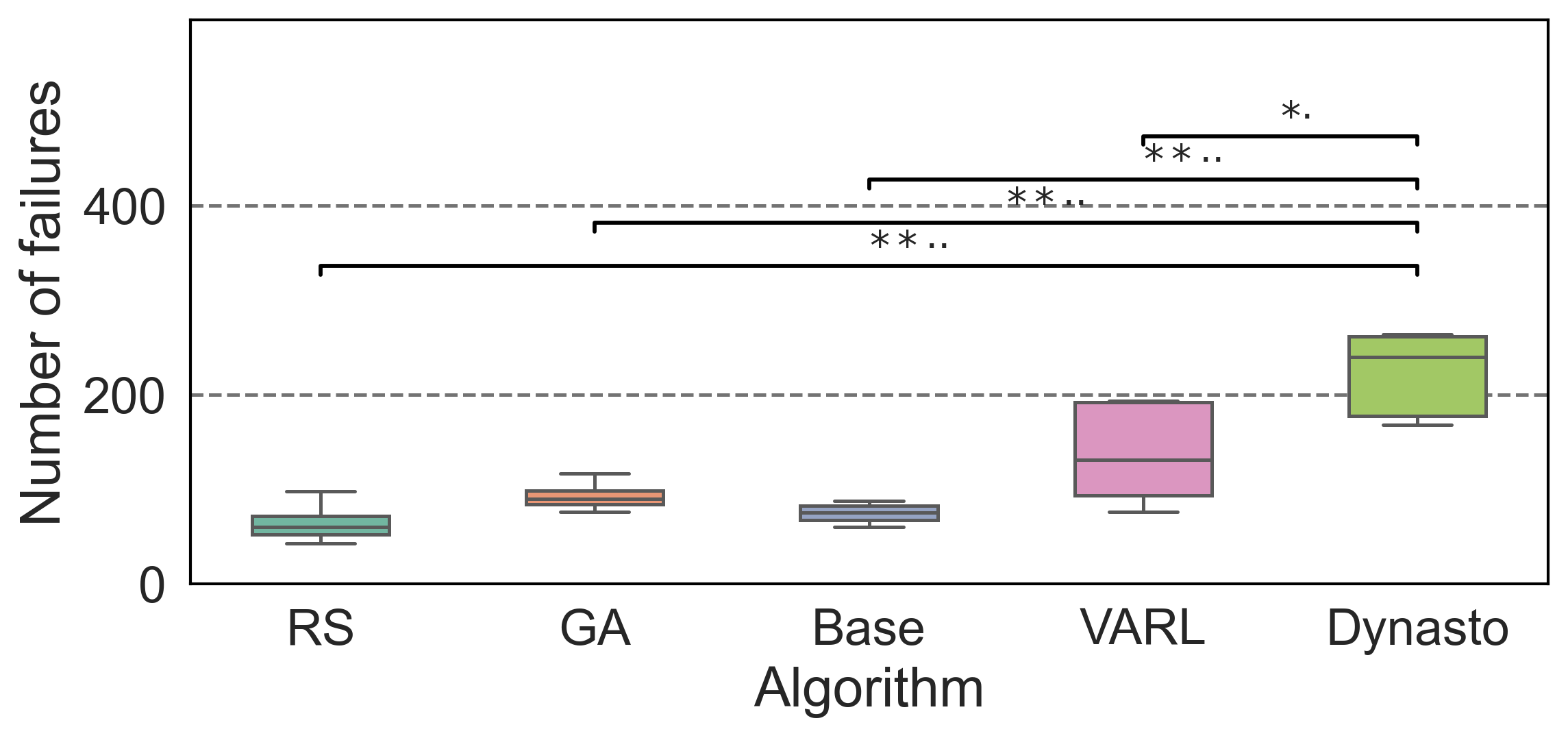}
\centering
\caption{Number of failures revealed by algorithm in UC2}
\label{fig:uc2_box}
\end{figure}

\noindent \textbf{Results.} 
Figures~\ref{fig:uc1_box} and~\ref{fig:uc2_box} report the cumulative number of the revealed valid failures out of 4000 evaluations budget. For both SUTs, GA consistently outperforms RS with a large effect size, confirming that guided search is more effective than naive exploration. GA also surpasses the Baseline~\cite{doreste2024adversarial}, despite its dynamic action selection, likely because the Baseline reward promotes all failures and is therefore less targeted for valid ones. Our validity-aware RL (VARL) adversary further improves over GA, with medium to large effect sizes on both SUTs, showing the benefit of context-dependent action selection with validity-aware rewards. Adding evolutionary search over initial conditions yields an additional gain: \textsc{Dynasto} finds up to 70\% more valid failures than DQN under the same budget, as reflected in  the final failure counts (Figures~\ref{fig:uc1_box} and~\ref{fig:uc2_box}).


\noindent \textbf{Conclusions.} \textsc{Dynasto} augments RL with a GA over initial conditions, uncovering up to 70\% more valid failures than the validity-aware DQN, which already outperforms the Baseline~\cite{doreste2024adversarial} under the same budget.

\subsubsection{RQ2 -- \textsc{Dynasto} Variants and Configurations}
\hfill \break
\noindent \textbf{Motivation.} In RQ2, we compare two integrations of RL with initial-condition search: \emph{co-evolution} (policy and initializations optimized jointly) and \emph{sequential} \textsc{Dynasto} (learn RL first, then replay traces to guide GA search).\\
\textbf{Method.} We compare five approaches across both SUTs:
\begin{itemize}
 \item \textit{VARL}: an adversarial validity-aware RL agent with no initial-condition search.
 \item \textit{VARL+GA}: co-evolution of RL with GA, same configuration as \textsc{Dynasto(GA)}, except the chromosome omits the failure-trace selection gene.
 \item \textit{VARL+RS}: co-evolution of RL with Random.
 \item \textit{\textsc{Dynasto(GA)}}: our proposed \textsc{Dynasto}.
 \item \textit{\textsc{Dynasto(RS)}}: \textsc{Dynasto} with Random instead of GA.
\end{itemize}
 
\begin{figure}[h!]
\includegraphics[scale=0.35]{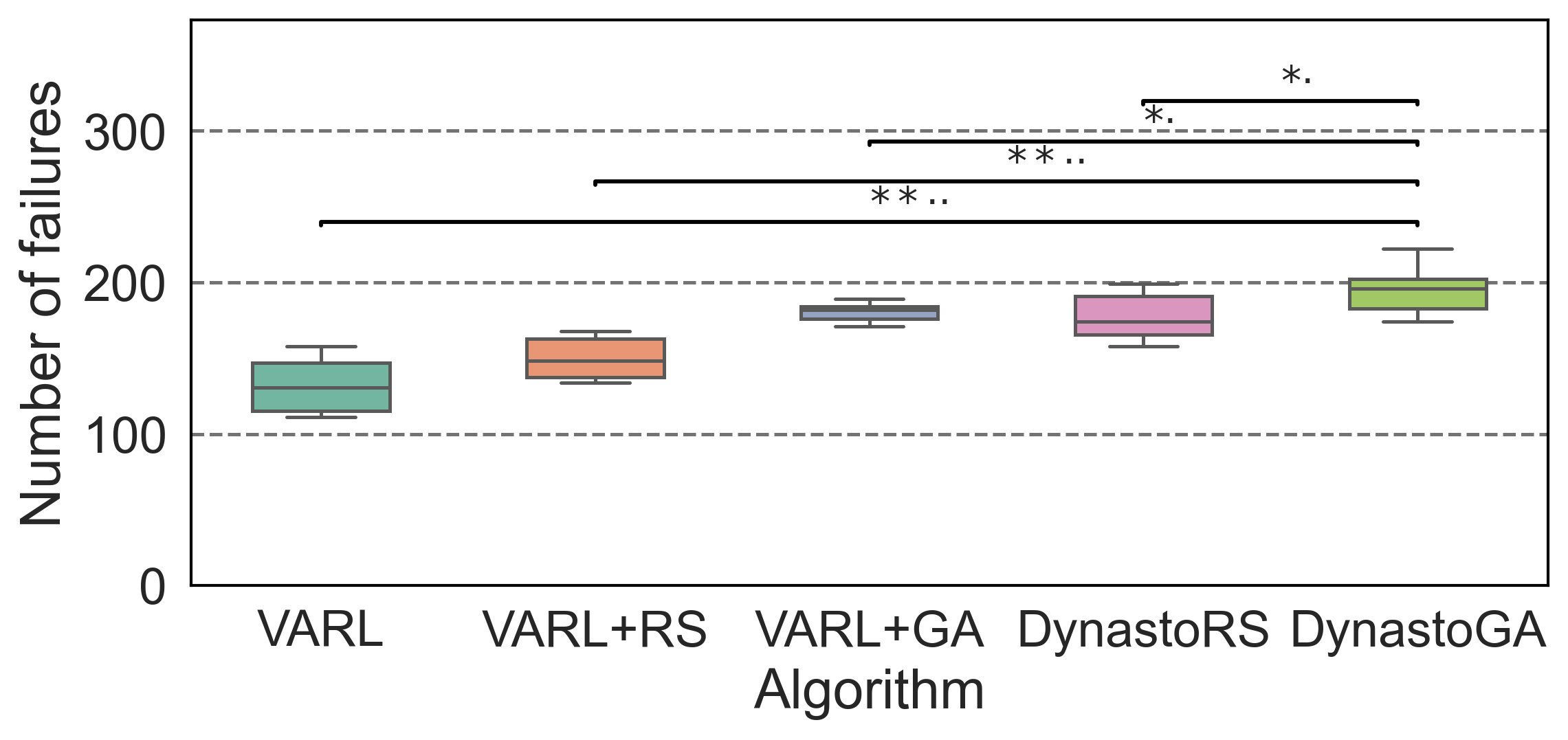}
\centering
\caption{Revealed failures count by configuration for UC1}
\label{fig:uc1-fail-rq2}
\end{figure}

\begin{figure}[h!]
\includegraphics[scale=0.35]{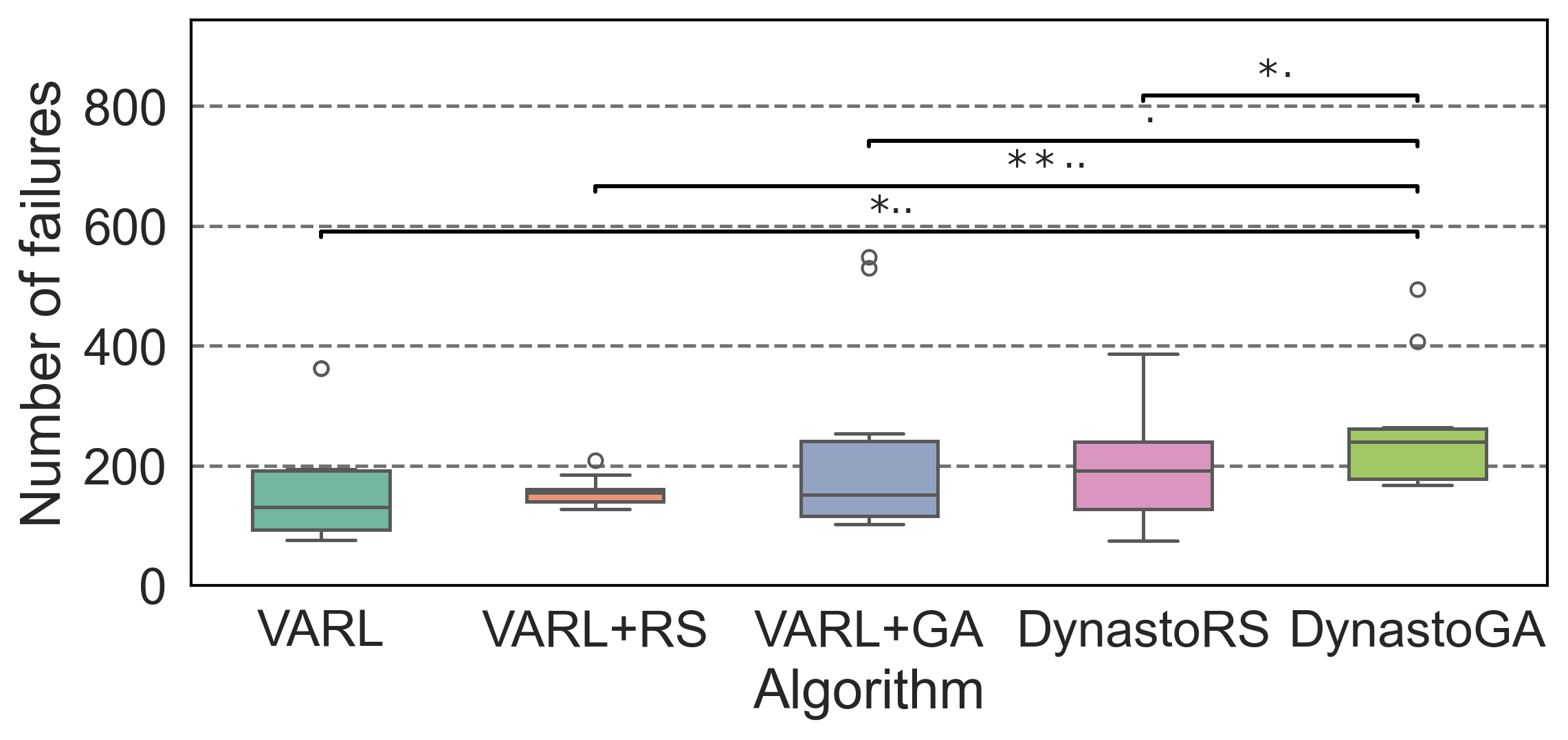}
\centering
\caption{Revealed failures count by configuration for UC2}
\label{fig:uc2-fail-rq2}
\end{figure}

\noindent \textbf{Results.} 
Figures~\ref{fig:uc1-fail-rq2}--\ref{fig:uc2-fail-rq2} report the cumulative count of discovered failures in UC1 and UC2, respectively. The statistical significance of the observed differences is summarized in the boxplots.


\noindent In UC1, adding initial-condition search to RL improves over the \textit{VARL} baseline in both co-evolutionary and sequential setups, with large effect sizes. Evolutionary variants (\textit{VARL+GA} and \textit{\textsc{Dynasto(GA)}}) also outperform their random-search counterparts (\textit{VARL+RS} and \textit{\textsc{Dynasto(RS)}}) with large effect sizes. Overall, \textit{\textsc{Dynasto(GA)}} is the strongest configuration, uncovering 11.4\% more failures than co-evolutionary \textit{VARL+GA}, with a medium effect size. In UC2, only \textit{\textsc{Dynasto(GA)}} improves over \textit{VARL} with a statistically significant, large effect size, revealing about 78\% more failures than both the baseline and the co-evolutionary setups. Within \textsc{Dynasto} variants, \textit{\textsc{Dynasto(GA)}} also outperforms \textit{\textsc{Dynasto(RS)}} with a medium effect size, yielding 31\% more failures on average.

\noindent \textbf{Conclusions.} \textsc{Dynasto} (\textit{dyntest\_ga}) outperforms all baselines on both SUTs. The co-evolutionary setup (\textit{ga\&rl}) can yield more failures in some runs but is less stable and more sensitive to initialization. \textsc{Dynasto} is more consistent because it optimizes one component at a time.

\subsection{RQ3 -- Failure Analysis and Diversity}
\begin{figure}[h]
\includegraphics[scale=0.33]{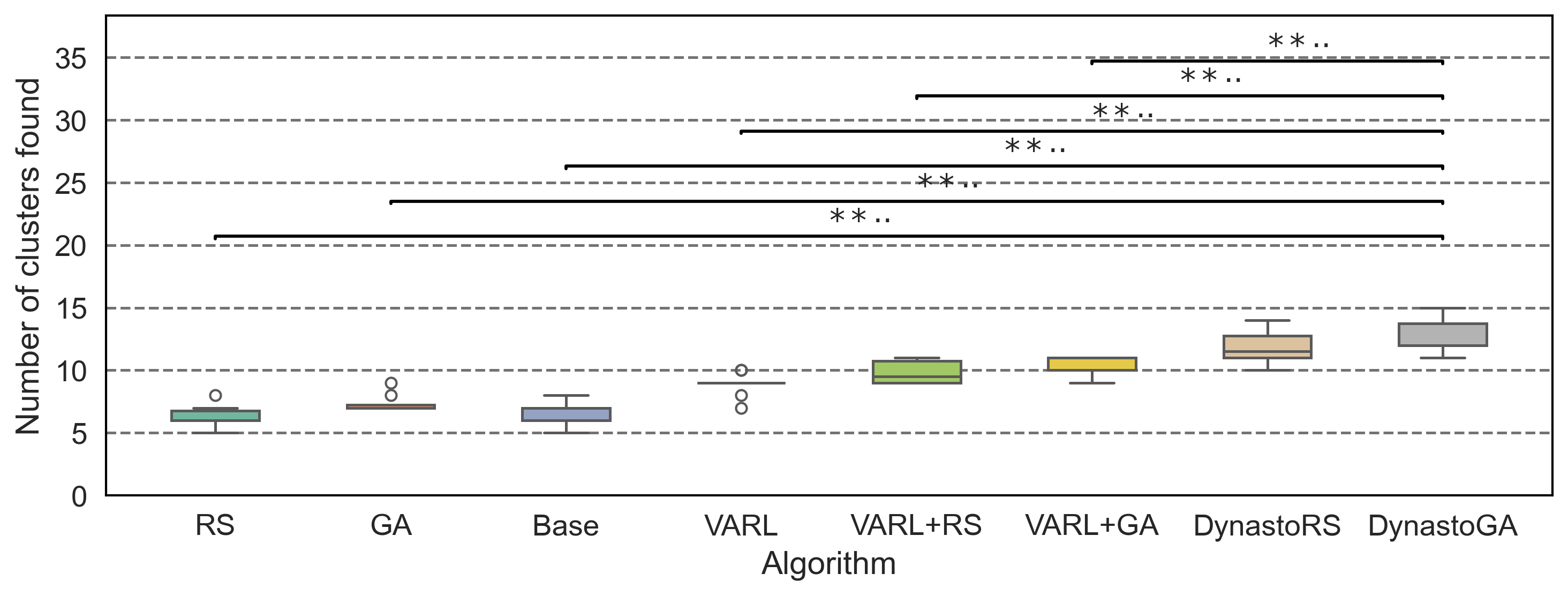}
\centering
\caption{Failure clusters count discovered by method for UC1}
\label{fig:uc1-cluster-rq3}
\end{figure}

\begin{figure}[h]
\includegraphics[scale=0.33]{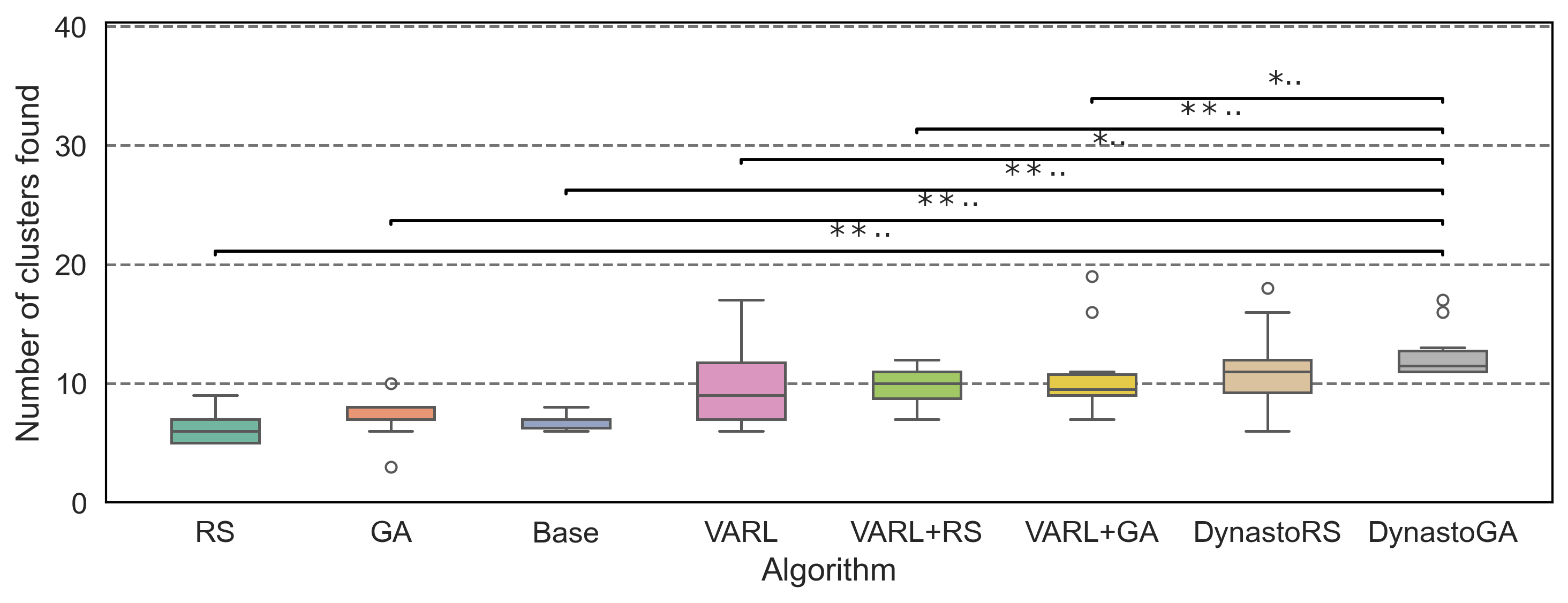}
\centering
\caption{Failure clusters count discovered by method for UC2}
\label{fig:uc2-cluster-rq3}
\end{figure}
\noindent \textbf{Motivation}. Unique-failure counts miss higher-level structure, so in this last RQ, we measure diversity by the number of failure clusters and characterize the dominant failure patterns.\\
\textbf{Method}. During scenario generation, we record all valid failures, remove duplicates, and cluster the remainder using the pipeline in Section~\ref{sec:post_hoc_analysis}. We then manually inspect the scenario videos within the clusters produced by \textsc{Dynasto(GA)}; all recordings are provided in the replication package.\\
\textbf{Results}. Figures~\ref{fig:uc1-cluster-rq3}--\ref{fig:uc2-cluster-rq3} report the cumulative number of clusters for UC1 and UC2, with statistical significance presented in the plots. The co-evolutionary setup that combines validity-aware RL (VARL) with initial-condition search improves over VARL without static-parameter search, revealing about 22\% more clusters with a large effect size. \textsc{Dynasto} outperforms all other methods with large effect sizes, uncovering up to 70\% more clusters. Comparing search strategies, both GA and RS discover new patterns; GA yields slightly more clusters on average, but the difference is not statistically significant, indicating exploration comparable to random sampling despite GA’s exploitation bias.

\begin{figure}[h]
    \centering
    \begin{subfigure}{0.48\linewidth}
        \centering
        \includegraphics[scale=0.21]{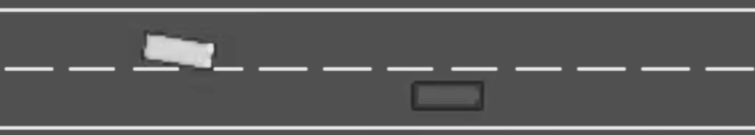}
        \caption{Ego vehicle doing an unsafe lane change}
    \end{subfigure}
    \hfill
    \begin{subfigure}{0.48\linewidth}
        \centering
        \includegraphics[scale=0.21]{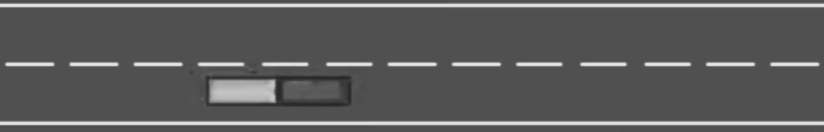}
        \caption{Collision with adversarial vehicle}
    \end{subfigure}
    \caption{Unsafe lane change failure pattern}
    \label{fig:fail1}
\end{figure}

\begin{figure}[h]
    \centering
    \begin{subfigure}{0.48\linewidth}
        \centering
        \includegraphics[scale=0.205]{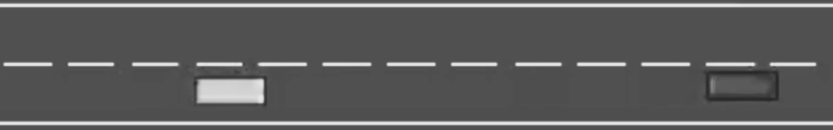}
        \caption{Adversarial vehicle braking in the same lane as the ego vehicle}
    \end{subfigure}
    \hfill
    \begin{subfigure}{0.48\linewidth}
        \centering
        \includegraphics[scale=0.20]{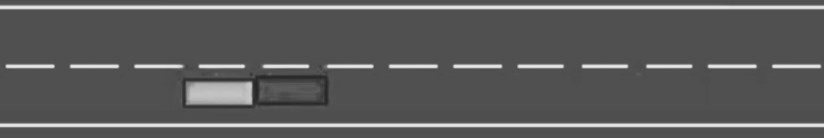}
        \caption{Ego vehicle failing to brake and adapt speed}
    \end{subfigure}
    \caption{Ego vehicle failing to brake}
    \label{fig:fail2}
\end{figure}

\begin{figure}[h]
    \centering
    \begin{subfigure}{0.48\linewidth}
        \centering
        \includegraphics[scale=0.201]{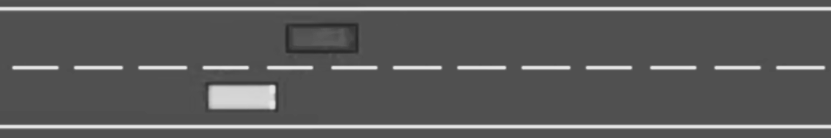}
        \caption{Ego vehicle performing an unsafe lane cut-in}
    \end{subfigure}
    \hfill
    \begin{subfigure}{0.48\linewidth}
        \centering
        \includegraphics[scale=0.21]{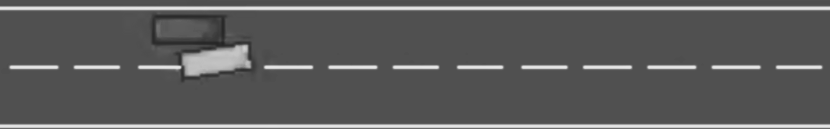}
        \caption{Collision with adversarial vehicle}
    \end{subfigure}
    \caption{Unsafe lane cut-in}
    \label{fig:fail3}
\end{figure}

\begin{figure}[h]
    \centering
    \begin{subfigure}{0.48\linewidth}
        \centering
        \includegraphics[scale=0.21]{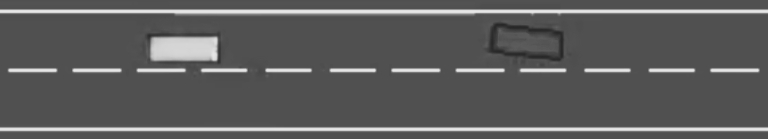}
        \caption{Adversary starting a lane change}
    \end{subfigure}
    \hfill
    \begin{subfigure}{0.48\linewidth}
        \centering
        \includegraphics[scale=0.21]{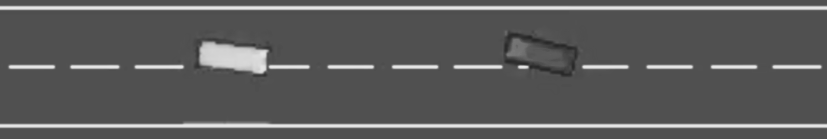}
        \caption{Ego vehicle performing the lane change after the adversary}
    \end{subfigure}

    \vspace{0.6em}

    \begin{subfigure}{0.6\linewidth}
        \centering
        \includegraphics[scale=0.225]{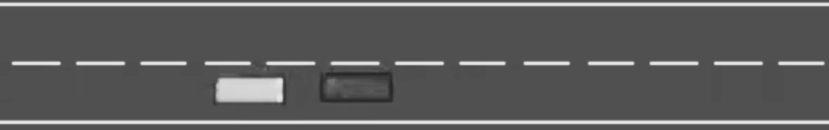}
        \caption{Ego vehicle colliding with the adversary}
    \end{subfigure}

    \caption{Ego vehicle following the adversary trajectory}
    \label{fig:fail4}
\end{figure}

\begin{figure}[h!]
    \centering
    \begin{subfigure}{0.48\linewidth}
        \centering
        \includegraphics[scale=0.21]{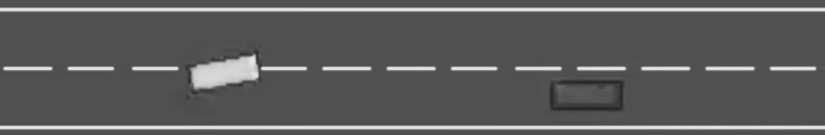}
        \caption{Ego vehicle avoiding the adversary}
    \end{subfigure}
    \hfill
    \begin{subfigure}{0.48\linewidth}
        \centering
        \includegraphics[scale=0.21]{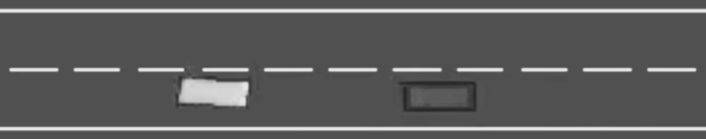}
        \caption{Ego vehicle performing an unsafe lane change}
    \end{subfigure}

    \vspace{0.6em}

    \begin{subfigure}{0.6\linewidth}
        \centering
        \includegraphics[scale=0.215]{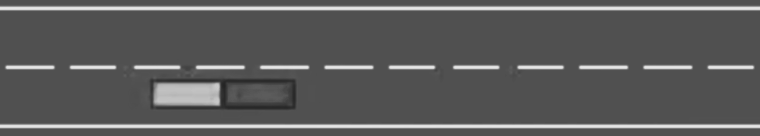}
        \caption{Collision with the adversarial vehicle}
    \end{subfigure}

    \caption{Unsafe lane change failure pattern}
    \label{fig:fail5}
\end{figure}

Next, we manually inspected representative clusters produced by \textsc{Dynasto} for both SUTs. 
Overall, the dominant patterns fall into two categories: unsafe lateral maneuvers and inadequate longitudinal response. Specifically, we observe (i) unsafe lane changes and side collisions when the ego (white rectangle) initiates a lateral maneuver with insufficient clearance to a nearby vehicle (black rectangle) (Figure \ref{fig:fail3}), (ii) rear-end collisions where the ego fails to brake when a lead vehicle slows down (Figure \ref{fig:fail2}), and (iii) unsafe cut-ins where the ego merges with insufficient headway, causing the following vehicle to collide with its rear (Figures \ref{fig:fail1}, \ref{fig:fail4}, \ref{fig:fail5}). We also observe a following-the-adversary pattern where the ego mirrors the adversary’s lane change too closely and collides due to insufficient lateral separation. Finally, we identify one pattern unique to SUT2 in which the ego first adjusts its lane position to avoid the adversary, then performs an unsafe lane change while the adversary remains adjacent, leading to a lateral collision.

\noindent \textbf{Conclusions.} \textsc{Dynasto} finds about 20\% more clusters than the co-evolutionary setup and 50\% more than the DQN baseline across both SUTs. Both GA- and random-search variants are effective, with GA yielding about 10\% more clusters on average. Clustering also compresses roughly 200 unique \textsc{Dynasto} failures into about 12 representative patterns per SUT, and manual inspection confirms these failures are valid and driven by unsafe ego behavior, revealing concrete weaknesses in decision-making and control.
\subsection{Data availability.}
Replication package including the trained models, results, and implementation scripts are publicly available at: \url{https://figshare.com/s/2a193e5eeff51f9b5b27}
\section{Related works} 
\label{sec:related}
In the section we review the existing works on ADS simulation-based testing and position our paper with the related work.\\
\textbf{Static parameter optimization.} 
A large body of search-based test generation methods for autonomous driving systems (ADS) focuses on optimizing \emph{static initial parameters}. Representative approaches evolve the initial positions of agents~\cite{abdessalem2018testing, abdessalem2018testing2, kluck2019genetic}, pedestrian initial states~\cite{tuncali2018simulation}, or ego-vehicle destinations, waypoints, and environmental conditions such as weather~\cite{ebadi2021efficient}. More recent work extends this line of research by optimizing predefined adversarial trajectories, maneuvers, vehicle speeds at discrete time instants, and obstacle placement~\cite{huai2023doppelganger, huai2023scenorita}. However, these approaches remain \emph{offline} in nature: adversarial behaviors are fixed prior to execution and do not adapt to the actions of other traffic participants during scenario rollout.

\textbf{Dynamic behavior generation.}
Another line of work employs online optimization techniques, most notably reinforcement learning (RL), to synthesize adversarial behaviors for ADS testing. Prior studies use RL to generate adversarial interactions by co-training ego vehicles and adversaries~\cite{chen2021adversarial, doreste2024adversarial}, modeling pedestrian disturbances~\cite{karunakaran2020efficient}, or framing simulator disturbances as RL actions in adaptive stress testing (AST)~\cite{lee2015adaptive, delecki2022we, lu2022learning}. Subsequent extensions consider richer, time-varying environments and multi-objective optimization~\cite{haq2023many}, or train separate RL policies for individual scenario participants~\cite{doreste2024adversarial}. While effective at producing interactive behaviors, these approaches do not systematically explore variations in static initial conditions nor incorporate scenario validity constraints into adversarial reward shaping, which can lead to unrealistic or overly aggressive behaviors.

\textbf{Failure analysis.} A limited number of works explicitly incorporate scenario validity rules~\cite{li2020av, li2024viohawk}. However, these approaches are predominantly fuzzing-based and do not involve dynamic decision making. In such methods, validity rules are typically enforced as search constraints rather than being treated as explicit optimization objectives.

\textbf{Positioning.} \textsc{Dynasto} separates dynamic and static optimization. Step~1 learns a single validity-aware adversarial policy with RL, guided by temporal-logic constraints. Step~2 searches over initial conditions with a GA while replaying the learned adversarial behavior, avoiding retraining multiple RL agents and systematically exposing failures that emerge from initialization. \textsc{Dynasto} further clusters failure traces to support structured analysis of distinct failure modes. 

\section{Threats to Validity}
\label{sec:threats}
We discuss threats to validity and mitigation measures.\\
\textbf{Internal validity.} All methods share the same well established simulator, action space, evaluation budget (4000 test executions), and validity criteria, but optimizers use different hyperparameters and convergence behavior, which can affect performance. We mitigate this by repeated runs with statistical testing and effect sizes  and by reusing components (environment interface and STL-based labeling) across methods.\\
\textbf{External validity.} Our evaluation is limited to 2D HighwayEnv simulation environment and 2-vehicle scenarios. To partially mitigate this threat we conducted experiments with different systems under test. \\ 
\textbf{Construct Validity.}
To mitigate threats to construct validity, we evaluate the proposed metrics on controlled inputs. Specifically, we calibrate the STL-based validity rules using a manually annotated set of 50 scenarios, tune the clustering algorithm on datasets with known cluster structure, and select the duplicate-removal threshold using a curated set of similar and dissimilar test scenarios. \\
\textbf{Conclusion Validity.}
Both reinforcement learning (RL) and genetic algorithms (GA) introduce stochasticity and run-to-run variability. We address this threat by repeating experiments multiple times, reporting statistical significance and effect sizes across runs.

 \section{Conclusions}
 \label{sec:conclusion}
Testing autonomous driving systems requires uncovering not only many failures but also a diverse set of behaviorally plausible, safety-relevant failure modes. We proposed \textsc{Dynasto}, a two-step scenario-generation approach that combines reinforcement learning for dynamic adversarial behavior with evolutionary search over static initial conditions. Our empirical evaluation on \textit{HighwayEnv} with two systems under test shows that a validity-aware RL adversary, used sequentially with GA search over initial conditions, improves failure discovery effectiveness, revealing over 22\% more failures patterns compared to baselines. Future work includes extending \textsc{Dynasto} to richer multi-agent ADS settings with higher-dimensional perception, refining notions of failure severity and coverage, evaluating the effectiveness of the approach and its failure validity awareness in more photorealistic, advanced autonomous driving simulators such as CARLA.

\bibliographystyle{IEEEtran}
\bibliography{IEEEabrv, refs}

\end{document}